# Statistical quantum operation

Sumiyoshi Abe[1,2,3], Yuichi Itto[1], and Mamoru Matsunaga[1]

[1]*Department of Physical Engineering, Mie University, Mie 514-8507, Japan*

[2]*Institut Supérieur des Matériaux et Mécaniques Avancés, 44 F. A. Bartholdi, 72000 Le Mans, France*

[3]*Inspire Institute Inc., McLean, Virginia 22101, USA*

**Abstract** A generic unital positive operator-valued measure (POVM), which transforms a given stationary pure state to an arbitrary statistical state with perfect decoherence, is presented. This allows one to operationally realize thermalization as a special case. The loss of information due to randomness generated by the operation is discussed by evaluating the entropy. Thermalization of the bipartite spin-1/2 system is discussed as an illustrative example.

## I. INTRODUCTION

The probabilistic element in quantum mechanics is more fundamental than that in statistical mechanics: the former is of inherence of nature, whereas the latter is concerned with ignorance or lack of knowledge. With this observation as well as the intrinsic discreteness in quantum mechanics, one may imagine that the foundations of statistical mechanics can be understood based on the quantum-mechanical principles, entirely or partially at least in the low-temperature regime. This issue actually has a long history involving the works of von Neumann, Schrödinger, Wigner, and others, and a number of efforts have been devoted to it in the literature (See Ref. [1] and the references quoted therein).

In recent years, this traditional issue has been revisited from various viewpoints, and some intriguing proposals have been made for understanding the quantum-mechanical origin of the emergence of (micro)canonical ensemble. They include setting an energy shell in a Hilbert space using a special type of interaction [2], the hidden gauge structure generating quantum entanglement [3], Hilbert-space average [4], and Levy's lemma for entangled systems [5]. There are also discussions about the ubiquity/typicality of the quantum (micro)canonical ensemble [6-8].

Statistical mechanics tells us that a system initially prepared in an arbitrary state tends to reach the equilibrium state in an irreversible manner, in spite of the fact that the underlying microscopic dynamics is reversible. The quantum-mechanical counterpart of the situation is that a given pure state becomes mixed through a process that cannot be described by any of unitary transformations. Here is the crux of relevance of the quantum measurement problems.

A traditional way to derive the canonical ensemble is to consider the objective system weakly interacting with the environment and then to eliminate the environmental degrees of freedom from the total system in equilibrium as the microcanonical ensemble. Quantum mechanically, the effect of the environment on the objective system through the weak interaction can be viewed as measurement, the process of which is nonunitary.

In this paper, we describe the transition of a given initial state of the objective system to a statistical state by replacing the effect of the environment with a quantum operation. We represent such an operation by a positive operator-valued measure (POVM) [9-12]. We construct an operator, *which maps a given stationary pure state to an arbitrary statistical state with perfect decoherence*. In this framework, the canonical ensemble is seen to be just a specific case. To examine the physical property of this operation, we study the loss of information due to randomness generated by it by evaluating the entropy. We also discuss thermalization of the bipartite spin-1/2 system as an example in order to illustrate the present scheme.

## II.　STATISTICAL QUANTUM OPERATION

Let us start our discussion with taking a quantum-mechanical system with a Hamiltonian, *H*, in *d* dimensions. *d* can arbitrarily be large, in general. The set of its energy eigenstates, $\{|u_n\rangle\}_{n=1, 2, ..., d}$, is assumed to form an orthonormal complete system. Therefore, it gives rise to a resolution of the identity operator: $I = \sum_{n=1}^{d} |u_n\rangle\langle u_n|$. The identity operator yields a quantum operation that keeps an arbitrary state unchanged. Our central idea here is to pick up two states, say $|u_m\rangle$ and $|u_n\rangle$, and to consider the transition between them. Thus, we employ the constructive way to present the following operator:

$$V_n^{(m)} = a_n \left( I - |u_m\rangle\langle u_m| - |u_n\rangle\langle u_n| + |u_m\rangle\langle u_n| + |u_n\rangle\langle u_m| \right), \tag{1}$$

where $a_n$ is a complex *c*-number and $n = 1, 2, ..., d$. Let $\rho$ be an arbitrary density matrix in *d* dimensions. The positive quantum operation on $\rho$ associated with the operator in Eq. (1) is defined by the following linear map:

$$\rho \rightarrow \Phi^{(m)}(\rho) \equiv \sum_{n=1}^{d} V_n^{(m)} \rho V_n^{(m)\dagger}. \tag{2}$$

By a straightforward calculation, one can ascertain that this operation possesses the following general properties:

(i) *trace-preserving*;

$$\sum_{n=1}^{d} V_n^{(m)\dagger} V_n^{(m)} = I \qquad (\forall m) \tag{3}$$

under the condition

$$\sum_{n=1}^{d} |a_n|^2 = 1, \tag{4}$$

from which $\mathrm{Tr}\,\Phi^{(m)}(\rho) = \mathrm{Tr}\,\rho = 1$ holds, and

(ii) *unital*;

$$\sum_{n=1}^{d} V_n^{(m)} V_n^{(m)\dagger} = I \qquad (\forall m) \tag{5}$$

which results from the fact that $V_n^{(m)}$ is a normal operator, i.e., $\left[V_n^{(m)}, V_n^{(m)\dagger}\right] = 0$, as well as the condition in Eq. (4). The property (ii) implies that the completely random state, $I/d$, is a fixed point of $\Phi^{(m)}$. Due to the condition in Eq. (4), $\{a_n\}_{n=1, 2, ..., d}$ is referred to here as "amplitude".

Thus, $V_n^{(m)}$ in Eq. (1) yields a unital POVM.

Unital operations play a distinguished role in quantum statistical mechanics and quantum measurement problems. This is because as follows. Let $f$ and $A$ be operator concave and Hermitian, respectively. Then, for a unital operation, $\Phi$, it follows [13] that $f(\Phi(A)) \geq \Phi(f(A))$. Putting $A = \rho$ and $f(\rho) = -\rho \ln \rho$, this leads to $S(\Phi(\rho)) \geq S(\rho)$ with the von Neumann entropy

$$S(\rho) = -\text{Tr}(\rho \ln \rho), \tag{6}$$

which means that $S(\rho)$ does not decrease under operations with unital POVM's, showing irreversibility.

Now, a point of crucial importance is that the quantum operation, $\Phi^{(m)}$, in Eq. (2) transforms the "initial" pure stationary state, $|u_m\rangle\langle u_m|$, as follows:

$$\Phi^{(m)}\left(|u_m\rangle\langle u_m|\right) = \sum_{n=1}^{d} |a_n|^2 |u_n\rangle\langle u_n|. \tag{7}$$

This "final" state on the right-hand side does not have off-diagonal elements and therefore perfect decoherence is realized. Recalling Eq. (4), it is seen to be an arbitrary statistical state. In addition, initial-state-dependence completely disappears.

In the special case when the amplitude is taken to satisfy

$$|a_n|^2 = \frac{e^{-\beta \varepsilon_n}}{Z(\beta)}, \tag{8}$$

$$Z(\beta) \equiv \sum_{n=1}^{d} e^{-\beta \varepsilon_n}, \tag{9}$$

with $\varepsilon_n$ being the $n$th eigenvalue of the Hamiltonian, Eq. (7) precisely becomes the canonical density matrix, $e^{-\beta H} / \text{Tr}\, e^{-\beta H}$, with inverse temperature, $\beta$.

Thus, we were successful in constructing a highly generic unital quantum operation,

which transforms a given stationary pure state to an arbitrary statistical state.

## III. LOSS OF INFORMATION BY THE QUANTUM OPERATION

Next, let us briefly discuss the loss of information due to randomness generated by the present operation by evaluating the entropy. The von Neumann entropy in Eq. (6) for the initial stationary pure state, $|u_m\rangle\langle u_m|$, vanishes. On the other hand, the value of the entropy for the final state is calculated to be

$$S_{\text{final}} \equiv S\left(\Phi^{(m)}(|u_m\rangle\langle u_m|)\right) = -\sum_{n=1}^{d} |a_n|^2 \ln |a_n|^2. \qquad (10)$$

This is the amount of the loss of information by the present POVM operation. In the case of thermalization in Eq. (8), $S_{\text{final}}$ is given by

$$S_{\text{final}} = \beta(U - F), \qquad (11)$$

where $U$ and $F$ are the internal energy, $\sum_{n=1}^{d} \varepsilon_n e^{-\beta \varepsilon_n} / Z(\beta)$, and Helmholtz free energy, $-\beta^{-1} \ln Z(\beta)$, in equilibrium, respectively.

## IV. EXAMPLE: BIPARTITE SPIN-1/2 SYSTEM

Finally, let us discuss thermalization of the bipartite spin-1/2 system as a simple and illustrative example. The Hamiltonian we employ here reads

$$H = -J\sigma_A \cdot \sigma_B, \quad (12)$$

where $\sigma$'s and $J$ are the Pauli matrices and a coupling constant, respectively. The dimensionality is $d = 4$. The energy eigenstates are given by

$$|u_1\rangle = \frac{1}{\sqrt{2}}\left(|\uparrow\rangle_A|\uparrow\rangle_B + |\downarrow\rangle_A|\downarrow\rangle_B\right), \quad |u_2\rangle = \frac{1}{\sqrt{2}}\left(|\uparrow\rangle_A|\uparrow\rangle_B - |\downarrow\rangle_A|\downarrow\rangle_B\right),$$

$$|u_3\rangle = \frac{1}{\sqrt{2}}\left(|\uparrow\rangle_A|\downarrow\rangle_B + |\downarrow\rangle_A|\uparrow\rangle_B\right), \quad |u_4\rangle = \frac{1}{\sqrt{2}}\left(|\uparrow\rangle_A|\downarrow\rangle_B - |\downarrow\rangle_A|\uparrow\rangle_B\right), \quad (13)$$

where $|\uparrow\rangle$ ($|\downarrow\rangle$) is the eigenstate of the $z$-component of the Pauli matrices with the eigenvalue, $+1$ ($-1$). $\{|u_n\rangle\}_{n=1,2,3,4}$ forms an orthonormal complete system termed the Bell basis. The corresponding energy eigenvalues are

$$\varepsilon_1 = \varepsilon_2 = \varepsilon_3 = -J, \quad \varepsilon_4 = 3J, \quad (14)$$

showing the existence of 3-fold degeneracy. Let us consider the quantum operation on the pure state, $|u_4\rangle\langle u_4|$. From the general scheme developed above, we obtain $\Phi^{(4)}(|u_4\rangle\langle u_4|) = \sum_{n=1}^{4} |a_n|^2 |u_n\rangle\langle u_n|$. In the case of thermalization, we have $|a_n|^2 = e^{-\beta\varepsilon_n}/Z(\beta)$, where $Z(\beta) = 3e^{\beta J} + e^{-3\beta J}$. Accordingly, the final state becomes $\Phi^{(4)}(|u_4\rangle\langle u_4|) = e^{-\beta H}/Z(\beta)$. We wish to make an additional remark. In this example,

the Hamiltonian can be recast in the form: $H = 4J|u_4\rangle\langle u_4| - JI$, which is essentially a projector Hamiltonian. The problem of quantum entanglement in this system has been studied from the view point of thermostatistics extensively in a recent work in Ref. [12].

## V. CONCLUSION

We have presented a generic unital POVM operation that transforms a given stationary pure state to an arbitrary statistical state. We have discussed the loss of information due to randomness generated by the operation. We have also illustrated the present scheme by considering thermalization of the bipartite spin-1/2 system as an example.

## ACKNOWLEDGMENT

S. A. was supported in part by a Grant-in-Aid for Scientific Research from the Japan Society for the Promotion of Science.

______________________________